\documentclass[12pt,preprint]{aastex}
\usepackage{emulateapj5}
\usepackage{apjfonts}

\newcommand{\h}{$H_0=70\, {\rm km\, s^{-1} Mpc^{-1}}$}

\newcommand{\epsi}{$g_{\rm k}\, $}
\newcommand{\epsikin}{$\epsilon_{\rm k}\, $}

\begin{document}

\def\sarc{$^{\prime\prime}\!\!.$}
\def\arcsec{$^{\prime\prime}$}
\def\arcmin{$^{\prime}$}
\def\degr{$^{\circ}$}
\def\seco{$^{\rm s}\!\!.$}
\def\ls{\lower 2pt \hbox{$\;\scriptscriptstyle \buildrel<\over\sim\;$}}
\def\gs{\lower 2pt \hbox{$\;\scriptscriptstyle \buildrel>\over\sim\;$}}
\def\lk{$L_{\rm KIN}$}

\title{The Optical-Radio Mapping: the kinetic efficiency of Radio-Loud AGNs}

\author{Francesco Shankar\altaffilmark{1}, Alfonso Cavaliere\altaffilmark{2}, Michele Cirasuolo\altaffilmark{3} and Laura Maraschi\altaffilmark{4}}
\altaffiltext{1}{Department of Astronomy,
 The Ohio State University, Columbus, OH 43210,
shankar@astronomy.ohio-state.edu} \altaffiltext{2}{Astrofisica, Dip.
Fisica Univ. "Tor Vergata", Via Ricerca Scientifica 1, 00133 Roma,
Italy} \altaffiltext{3}{SUPA Institute for Astronomy, University of
Edinburgh, Royal Observatory, Edinburgh EH9, 3HJ}
\altaffiltext{4}{INAF-Osservatorio Astronomico di Brera, via Brera
28, 20121 Milano, Italy}

%\begin{doublespace}

\begin{abstract}
We constrain the mean kinetic efficiency of radio-loud active
galactic nuclei by using an optically selected sample for which both
the optical and the radio luminosity functions (LFs) have been
determined; the former traces the bolometric luminosity $L$, while
the latter traces the kinetic power $L_{\rm k}$, empirically
correlated to the radio emission. Thus in terms of the ratio $g_{\rm
k}=L_{\rm k}/L$, we can convert the optical LF of the sample into a
radio one. This computed LF is shown to match the directly observed
LF for the same sample if $g_{\rm k}=0.10^{+0.05}_{-0.01}$ holds,
with a scatter $\sigma=0.38^{+0.04}_{-0.09}$ dex; with these values
we also match a number of independent correlations between $L_{\rm
k}$, $L$ and radio emission, that we derive through Monte Carlo
simulations. We proceed to translate the value of $g_{\rm k}$ into a
constraint on the kinetic efficiency for the production of radio
jets or winds, namely, $\epsilon_{\rm k}=L_{\rm k}/\dot{M}_{\rm
a}c^2\sim 0.01$ in terms of the rate $\dot{M}_{\rm a}$ of mass
accretion onto the central black hole. Then, on assuming that on
average the radio sources share the same kinetic efficiency, we
compute a solid lower limit of about $25\%$ on the contribution of
radio sources to the local black hole mass density.
\end{abstract}


\section{INTRODUCTION}
\label{sec|intro}

The origin of the radio-loudness of Active Galactic Nuclei (AGN)
still constitutes an open issue. There is evidence suggesting that
the formation of a relativistic jet or a fast wind (e.g., Blundell
\& Kuncic 2007) sustaining the radio emission is tightly related to
the mass of the central black hole (BH; e.g., Laor 2000). On the
other hand, the wide scatter observed between radio and optical
luminosities (e.g., Cirasuolo et al. 2003a) suggests that other
parameters such as the mass accretion rate onto the BH and possibly
its spin could play a significant role in determining when a galaxy
becomes radio-loud (e.g., Blandford et al. 1990, Ho 2002, Best et
al. 2005, Sikora et al. 2005 and references therein).

At one enticing extreme, Blandford \& Znajek (1977) proposed the
jets to be powered by the extraction of energy already accumulated
in a rotating BH. On the other hand, the spin, however important to
set the jet direction, may not provide the necessary power for
energizing the very luminous sources such as some Steep Spectrum
Radio sources (SSRs) and many Flat Spectrum Quasars (FSQs).

Alternative models (e.g., Livio et al. 1999) have proposed the
dominant fraction of the jet or wind kinetic power to be directly
linked to the rest-mass energy of the currently accreting matter, as
is the case for the radiative power. In any event, the kinetic power
must originate at some time from the accretion energy; thus a
constraint on the link between these two quantities is of key
importance to probe the still unsettled issues of the jet's
composition and production mechanisms (e.g., Celotti 2004).
%In particular one of the mechanisms proposed to explain the
%formation of the jets is the extraction of energy accumulated in a
%rotating BH (Blandford \& Znajek 1977). This implies a power
%proportional to the square of the intensity magnetic field threading
%the BH horizon and to the square of the angular momentum. However
%important to give the jet direction, the spin energy may not provide
%the necessary power to energize the very luminous sources such as
%the Steep Spectrum (SSR) and Flat Spectrum Quasars (FSQ) (Livio et
%al. 1999). Several theoretical models have proposed that a
%significant fraction of the jet kinetic power must be directly
%linked with the rest-mass energy of the accretion matter, as in the
%case of the radiated energy.

In this work we propose a simple but efficient and statistically
significant way to constrain the fraction of \emph{kinetic} to
accretion power. In the case of the radiative bolometric luminosity
$L$, the emitted power is written as $L=\epsilon \dot{M}_{\rm
a}c^2$, where $\dot{M}_{\rm a}$ is the current accretion rate of
matter onto the central massive BH and $\epsilon$ is the efficiency
to extract $L$ from $\dot{M}_{\rm a}$. In an analogous way we can
set the kinetic output as $L_{\rm k}=\epsilon_{\rm k}\dot{M}_{\rm
a}c^2=(\epsilon_{\rm k}/ \epsilon)\, L$. We assume that the AGN
kinetic output represents a fraction \epsi of the bolometric
luminosity, i.e.,
\begin{equation}
L_{\rm k}=g_{\rm k}L\, ;
    \label{eq|LkinL}
\end{equation}
this implies that $g_{\rm k}=\epsilon_{\rm k}/ \epsilon$ if the two
outputs draw from the same accretion flow. Thus one could constrain
$g_{\rm k}$ and in turn $\epsilon_{\rm k}$ by selecting a
statistically significant sample of radio-loud AGNs for which both
the radio and optical luminosities are known. The latter is in fact
a good tracer of the bolometric emission, given that AGN spectral
studies have provided over the years a reliable average bolometric
correction $C_B=L/L_B\nu_B\approx 10$, where the $B$-band luminosity
$L_B$ is in units of ${\rm erg\,s^{-1}Hz^{-1}}$ and $\nu_B=6.8\times
10^{14}\, {\rm Hz}$ (Elvis et al. 1994). In turn, significant radio
emission is always associated with kinetic power, and several papers
(referred to in \S~\ref{sec|Method}) provide and discuss empirical
correlations between $L_{\rm k}$ and the radio luminosity $L_{\rm
R}$.

Our aim is to \emph{statistically} derive the fraction of the
accreting rest mass energy that powers the jets or winds by
computing the average ratio $g_{\rm k}$, and then derive a lower
limit to the predicted local BH mass density associated with
radio-loud AGNs. In \S~\ref{sec|Method} we will present our method,
the results of which will be given in \S~\ref{sec|results}. In
\S~\ref{sec|soltan} we will describe in detail how to compute the
corresponding contribution to the BH mass density using the value of
\epsikin. In \S~\ref{sec|conclu} we will discuss our results and
give our conclusions.

In the following we will adopt the "concordance cosmology" (see
Spergel et al. 2007) with round parameters \h, $\Omega_m=0.3$ and
$\Omega_{\Lambda}=0.7$.

\section{The method}
\label{sec|Method}

In this \S$\, $we describe how we statistically constrain the ratio
between kinetic and bolometric luminosities $g_{\rm k}$. Our
database will be constituted by the optical and the radio luminosity
functions (LFs) derived by Cirasuolo, Magliocchetti \& Celotti
(2005; henceforth CMC) for the same sample of optically selected
radio-loud AGNs; this is collected by cross-correlating the 2-degree
Field Quasar Redshift survey (2dF) with the Faint Images Radio Sky
at Twenty cm (FIRST). The sample spans a considerable range in
magnitude $-24\lesssim M_B\lesssim -28$ and redshift $0.8\lesssim z
\lesssim 2.2$. In the following we will start from their optical
LF\footnote{Here we point out that the correct values of the CMC
double power-law LF should read: $\alpha=0.228$, $\beta=2.37$ for
the slopes, $\Phi_0=6.1\times10^{-8}\,\, {\rm Mpc^{-3}}$ for the
normalization and $M^*(z)=M^*(0)-2.5(k_1 z+k_2 z^2)$, with
$k_1=2.24$, $k_2=-0.696$, for the break luminosity.} with parameters
adjusted to our cosmology.

We then convert the adopted optical LF to a bolometric one by using
the average value $C_B=10.4$ with a log-normal scatter of 0.1 dex
around the mean. We have checked that our results are not sensitive
to the precise value of $C_B$, as very similar conclusions are found
also on using $C_B\approx 8$ (e.g., Marconi et al. 2004, Richards et
al. 2006). The direct link between optical and radio luminosities
has been already extensively studied by CMC (and references therein)
who find that a correlation between these two quantities exists,
although with a wide scatter.

In this work we are mainly focused on constraining the average
kinetic output of radio-loud AGNs. Several studies have been able to
empirically define the relation between radio emissions and kinetic
outputs, showing that on average the former are associated to high
levels of the latter (e.g., Rawling and Saunders 1991). After
expressing the bolometric LF in terms of kinetic luminosities via
Eq.~(\ref{eq|LkinL}), we convert to radio powers on using the
relation of Willott et al. (1999)
\begin{equation}
L_{\rm k}=3\times 10^{45} f^{3/2}L_{151}^{6/7} {\rm \,\, erg\,
s^{-1}}\, ,
%L_{151}=f^{-7/4}\left(\frac{L_{\rm k}}{3\times
%10^{45}}\right)^{7/6}\, ,
    \label{eq|Willott}
\end{equation}
where $L_{151}$ is the observed radio luminosity in units of
$10^{28}\, {\rm W\, Hz^{-1}\, sr^{-1}}$ at 151 MHz.

In Eq.~(\ref{eq|Willott}) $L_{\rm k}$ was empirically measured on
tracing the integrated $pdV$ work done by radio AGNs in excavating
the cavities observed in the hot gaseous medium around them. The
relation calibrated by Willott et al. (1999) relies therefore on
specific assumptions on how to link the age of the source to its
kinetic power. Following Hardcastle et al. (2007), the factor $f$ is
introduced in Eq.~(\ref{eq|Willott}) to account for systematic
underestimates of the true jet power that this technique may induce;
for example, the quantity $f$ also includes the coupling efficiency
between the AGN kinetic output and the surrounding medium. In the
following we will use (and discuss in \S~\ref{sec|conclu}) the
average value $f=15$, as measured for a sample of Faranoff-Rayleigh
I and II radio galaxies (Hardcastle et al. 2007 and references
therein).

Note that both $L_k$ and $L_{151}$ in Eq.~(\ref{eq|Willott}) are
calibrated on samples mainly composed of radio galaxies; however, if
the unification model is to hold on average for all radio sources
(see Urry \& Padovani 1995), AGNs and radio galaxies are similar
physical systems only observed at different angles; so we will take
Eq.~(\ref{eq|Willott}) to hold also for our sample of radio-loud
AGNs. Throughout we will adopt an average AGN radio spectral slope
$\alpha_R=0.7$, typical of the bright Steep Spectrum radio sources
sampled by CMC (see also references therein). A Gaussian scatter
with variance $\sigma$ is adopted around the mean of both
Eqs.~(\ref{eq|LkinL}) and ~(\ref{eq|Willott}).

Thus we obtain a radio LF which depends on only two free parameters,
\epsi and $\sigma$, once $f$ is fixed. The values of \epsi and
$\sigma$ are then constrained through a $\chi^2$ analysis by
matching our result to the radio LF independently measured by CMC.
%This value is also close to the average value of $f\sim 10$
%derived by Maraschi (2004) for a sample of blazars.

\section{RESULTS OF OPTICAL-RADIO MAPPING}
\label{sec|results}

\begin{figure}
\epsscale{0.8} \plotone{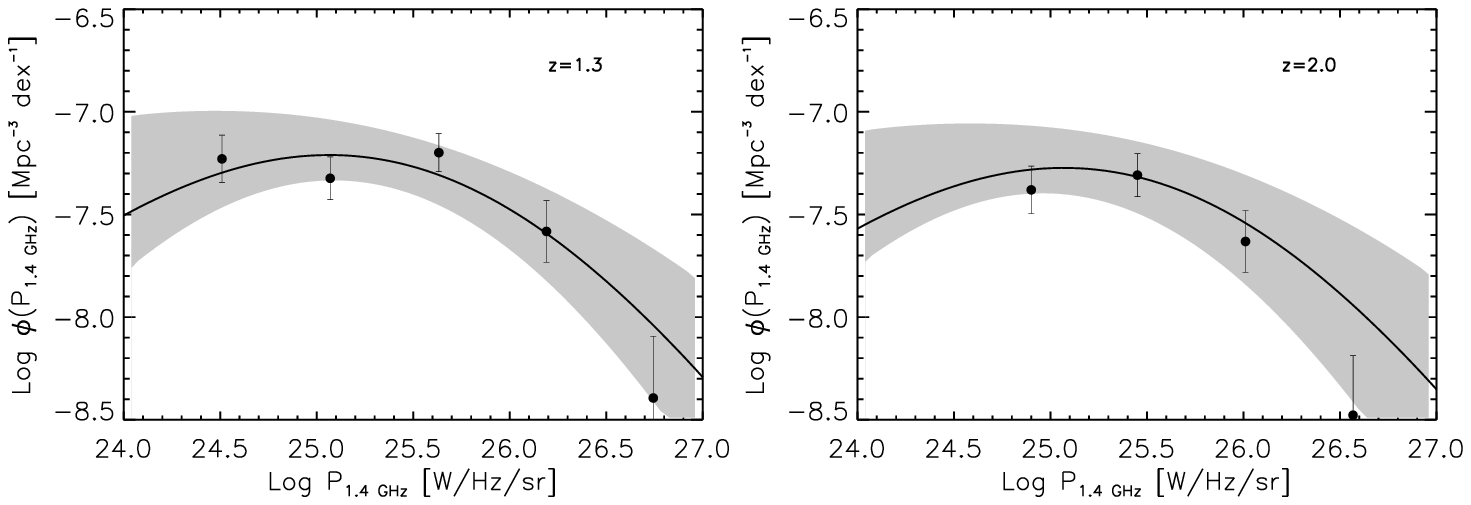}\caption{\small Best-fit radio LF
(\emph{solid} line) and its propagated $1-\sigma$ uncertainty region
(\emph{grey} area) obtained by converting the optical LF from CMC
with the use of Eqs.~(\ref{eq|LkinL}) and ~(\ref{eq|Willott}). The
result is presented for two redshifts bins as labeled, and is
compared with the data on the radio LF empirically derived by CMC
(\emph{solid} circles) for the same sample.} \label{fig|Phiradio}
\end{figure}

Our computed radio LF, shown as a solid line in
Fig.~\ref{fig|Phiradio}, fits the CMC data on the empirical radio LF
(solid points) when
\begin{eqnarray}
g_{\rm k}=0.10^{+0.05}_{-0.01} \nonumber\\
\sigma=0.38^{+0.04}_{-0.09}\, . \label{eq|gkin}
\end{eqnarray}
The grey area in Fig.~\ref{fig|Phiradio} shows the propagated
$1\sigma$ uncertainty from the optical LF. The match is good at all
luminosities and redshifts in the sample using the same value for
$g_{\rm k}$ (shown are the results for $z=1.3$ and $z=2$).
%Eq.~(\ref{eq|gkin}) implies a relatively small amount of power
%carried by the jet relative to radiative emission; if the observed
%jet power were close to the true kinetic power, i.e., if $f\sim 1$
%in Eq.~(\ref{eq|Willott}), then $g_{\rm k}$ would be even
%smaller. %The
%error-bars on \epsi are derived by matching the transformed optical
%LF with the $1\sigma$ uncertainties in the CMC radio data.
%(all three in units of $erg\, s^{-1}$)

\begin{figure}
\epsscale{0.8} \plotone{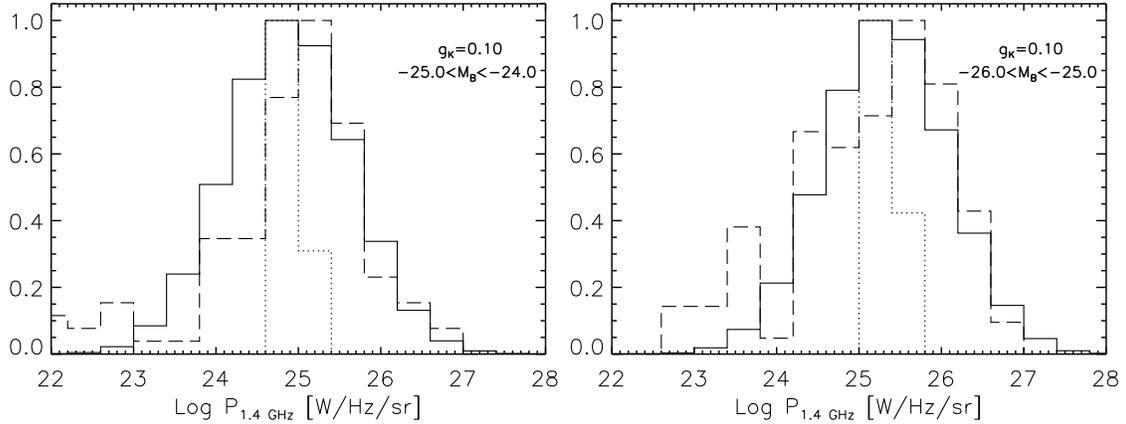}\caption{\small Comparisons between
the normalized distributions of sources as a function of radio power
corresponding to a given bin in optical luminosity, as labeled. The
\emph{dashed} line in each panel corresponds to the sample by CMC,
while the \emph{solid} line is the mean computed from our
simulations with scatter included. The \emph{dotted} line is the
distribution expected when no scatter is considered.}
\label{fig|simulations}
\end{figure}

We note that on adopting a Gaussian scatter we are able to reproduce
the mean and extent of the observed scatter around it, within a
given bin of optical luminosity. Fig.~\ref{fig|simulations} shows
our results (solid lines) for $-24\lesssim M_B\lesssim -25$ and
$-25\lesssim M_B\lesssim -26$ compared to the data collected by CMC
(dashed lines). Neglecting scatter in the relations (dotted lines)
would instead lead to a severe underestimate of the significant
spread observed in the data. %We also note that incompleteness
%effects are accounted for in the optical and radio LFs by CMC.

\begin{figure}
\epsscale{0.8} \plotone{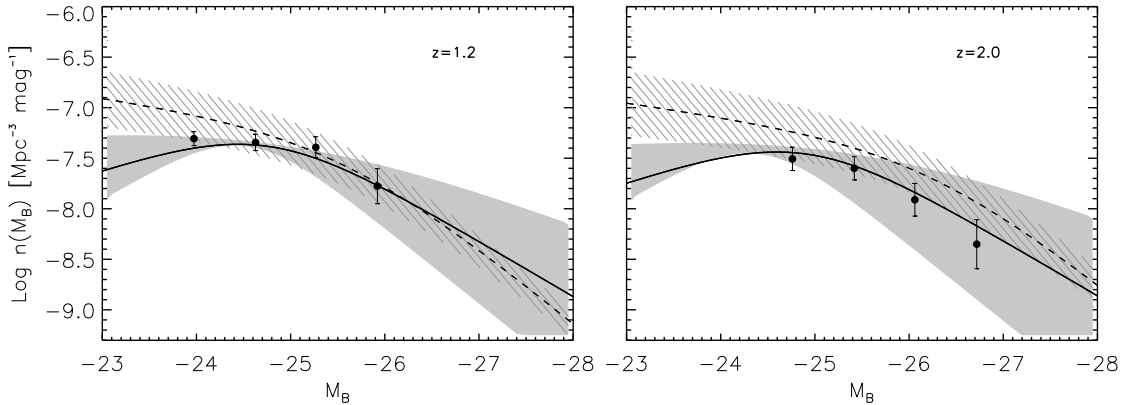}\caption{\small The \emph{solid}
line (with its 1-$\sigma$ uncertainty region shown with a grey area)
represents the fit to the optical LF, derived by CMC for their
sample of optical-radio sources in the 2dF-FIRST catalogues; the
solid points with error bars represent their original measurements.
The \emph{dashed} line and the striped area represent the best-fit
and $1-\sigma$ uncertainty region of the optical LF by Richards et
al. (2005), representative of the whole optical AGN population,
corrected for the luminosity and redshift-dependent fraction of
radio-loud AGNs derived by Jiang et al. (2007). These two
independent evaluations are compared at $z=1.2$ and $z=2.0$, as
indicated in the two panels; note the good agreement.}
\label{fig|PhiOpt}
\end{figure}

We have also performed other checks and tests on our results. To
look for possible biases in the optical selection of radio sources,
we have compared our determination of the optical LF with the recent
results by Jiang et al. (2007). These authors have cross-correlated
a large sample of optically selected quasars from SDSS with FIRST,
determining the radio-loud fraction of quasars as a function of
redshift and luminosity. By multiplying the LF by Richards et al.
(2005), representative of the whole population of optical AGNs, by
the Jiang et al. (2007) fraction of radio sources as a function of
optical luminosity, we can evaluate the radio-optical LF expected
from the SDSS data; in Fig.~\ref{fig|PhiOpt} we show this with a
dashed line in two reference redshift bins at $z=1.3$ and $z=2$.
Also shown with a striped area is the $1-\sigma$ uncertainty region
derived from the uncertainty in the Jiang et al. (2007) radio-loud
fraction. On the same plots we compare the completeness-corrected
evaluations by CMC for their radio-optical sample; we show their
data points as solid filled circles with error bars, together with
their best-fit estimate represented with a solid line. The good
match between the two evaluations supports the absence of any
specific bias in our adopted LFs, at least for $M_B\lesssim -24.5$
where most of the sample sources reside (cfr.
Fig.~\ref{fig|simulations}).

Through Monte Carlo sampling we then extract a sample of sources
from the optical LF with a given $L_B$ \footnote{All simulation
results presented in the following are provided by averages over
1000 realizations, each one with 10000 points.}, from which we
compute $L$, $L_{\rm k}$ and $L_{\rm R}$ by using
Eqs.~(\ref{eq|LkinL}) and (\ref{eq|Willott}) with their scatters
having fixed $g_{\rm k}$ to the value given in Eq.~(\ref{eq|gkin}).
In this way we build simulated distributions and correlations among
these observables, which we compare with a number of different data
sets.

%We sample bins of
%luminosities below and above the peak in the optical LF as measured
%by CMC, finding very good correspondence with the data.
%We cannot compare with the very low/high luminosity ends of the LF,
%due to poor statistics in the data. Moreover the comparison with CMC
%shows that our simple prescription of a Gaussian scatter can account
%for most of all the possible, if any, beaming effects in the sample.
%
%We have also simulated beaming effects following Urry \& Padovani
%(1991), which however produce much skewer distributions than those
%observed.

\begin{figure}
\epsscale{0.8} \plotone{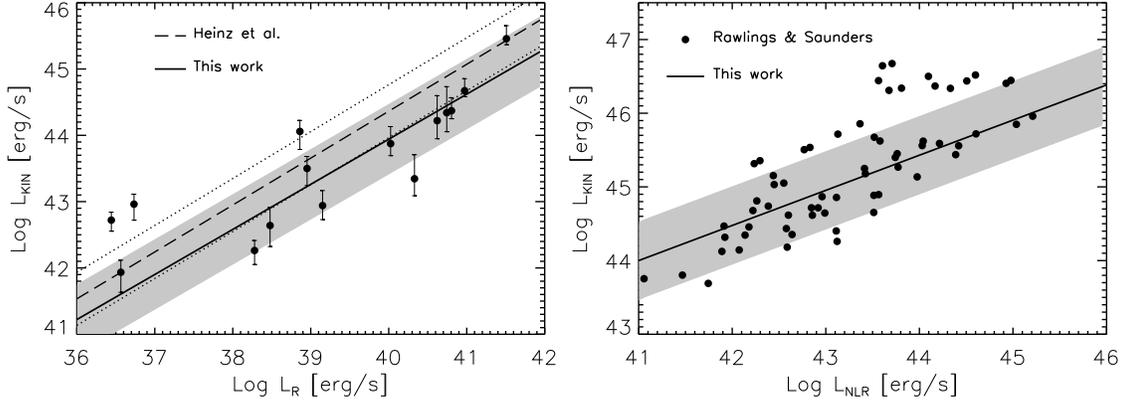}\caption{\small Comparison between
simulations and data. \emph{Left} panel: the results from our
best-fit model are compared with the set of data (filled circles)
presented in Heinz et al. (2007) and with their model
(\emph{long-dashed} line and \emph{dotted} lines). \emph{Right}
panel: our simulated best-fit relation (\emph{solid} line with
dispersion shown by the \emph{grey} area) is compared with data
derived by Rawling \& Saunders (1991). All data have been converted
to our adopted cosmology.} \label{fig|relations}
\end{figure}

%We also show in  that our simulated kinetic luminosities compare
%well with other sets of independent data.
Fig.~\ref{fig|relations} (left panel) shows our simulated $L_{\rm
R}-L_{\rm k}$ correlation, which is equivalent in slope, and
slightly lower in normalization but still compatible, relative to
the one derived by Heinz, Merloni \& Schwab (2007), and shown by the
long-dashed line with $1\sigma$ uncertainties shown as dotted lines.
Note that their relation is calibrated on a sample of FSQs, which
could be more beamed and/or energetic than our SSR dominated sample
(see also \S~\ref{sec|conclu}). Also shown with solid points and
error bars are the sets of data by B\^{i}rzan et al. (2004) and
Allen et al. (2006).

As Eq.~(\ref{eq|Willott}) might appear to be model-dependent, in the
right panel of Fig.~\ref{fig|relations} we also show how our results
on $L_{\rm k}$ relate to the narrow emission lines luminosities
$L_{\rm NLR}$\footnote{We use the conversions $L_{\rm NLR}=3(3L_{\rm
OII}+1.5L_{\rm OIII})$, $L_{\rm OII}=L_{\rm OIII}$ and $L_{\rm
OII}=L/5\times 10^3$ with scatter equal to $\sigma$ (e.g Willott et
al. 1999).}, and compare with the data points collected by Rawlings
\& Saunders (1991). The good agreement between our results and
theirs, which were derived under different assumptions from Willott
et al.
(1999), is encouraging. %The good agreement between their findings on the
\section{THE KINETIC CONTRIBUTION TO THE BH MASS FUNCTION}
\label{sec|soltan}

Constraining the total kinetic power from massive BHs is of key
importance for several models of galaxy evolution (e.g., Croton et
al. 2006, Cavaliere \& Lapi 2007) which require a significant amount
of kinetic feedback from the massive BHs to prevent catastrophic
cooling in the cores of groups and clusters of galaxies and limit
the formation of too massive galaxies in the local universe.

Thus in this \S$\, $we use our above results to evaluate the total
kinetic energy associated to SSRs. As discussed later in
\S~\ref{sec|conclu}, we shall not include additional contributions
to the kinetic integrated AGN emission from other other kinds of
AGNs, such as FSQs or radiatively inefficient sources, whose
modeling is uncertain. Rather, our aim is to provide a firm
\emph{lower} limit to the actual contribution of radio-kinetic AGNs
to the local mass density in black holes.

To compute this, we reverse the classic So{\l}tan (1982) approach,
which consists of estimating the average radiative efficiency
$\epsilon$ by dividing the integrated energy density observed from
AGNs by the local mass density in relic BHs. After computing the
integrated kinetic energy from SSRs, we divide it by our best-fit
value for the kinetic efficiency \epsikin, and time-integrate to
derive the local BH mass function expected from kinetically active
SSRs. By comparing the latter with the local mass function of the
whole BH population, we derive a lower limit to the fraction of the
relic BHs which have been kinetically active in their past.

We start from considering that during an accretion episode onto the
central BH a fraction of the accreting mass energy is released as
radiative and/or kinetic power. In \S~\ref{sec|results} we evaluated
the ratio \epsi to be of order $10^{-1}$, which in turn implies a
kinetic efficiency of \epsikin$\sim 10^{-2}$, if the sources in the
CMC sample possess on average the same radiative efficiency
$\epsilon\sim 0.1$ typical of AGNs (e.g., So{\l}tan 1982). Here we
make the specific assumption that this value of $\epsilon_k$ is
common to all Steep Spectrum kinetically-loud AGNs, irrespective of
their radiative emission. The total energy extracted reads
\begin{equation}
L_{\rm tot}=L+L_{\rm k}=(\epsilon+\epsilon_{\rm k})\,
\dot{M}_{a}c^2\, .
    \label{eq|LMdot}
\end{equation}
Therefore the total mass accreted onto the central BH is
\begin{equation}
\dot{M}_{BH}=(1-\epsilon-\epsilon_{\rm k})\, \dot{M}_{a}\, ,
\label{eq|MdotBH}
\end{equation}
which converts to
\begin{equation}
\dot{M}_{BH}=\frac{(1-\epsilon-\epsilon_{\rm k})\, L_{\rm
k}}{\epsilon_{\rm k}c^2}\,  \label{eq|MdotBHLkin}
\end{equation}
on using Eq.~(\ref{eq|LMdot}).
%, or also
%\begin{equation}
%\dot{M}_{BH}=\frac{(1-\epsilon-\epsilon_{\rm k})L}{\epsilon c^2}\,
%. \label{eq|MdotBHLrad}
%\end{equation}

By integrating the last equation over time and luminosity, and
equating it to the local mass density $\rho_{\rm BH}$ in relic BHs,
we obtain the kinetic ``So{\l}tan-type'' argument
\begin{equation}
\rho_{\rm BH}=\int dz\frac{dt}{dz} \int d\log L_{\rm k}\,
\Phi(L_{\rm k},z)\, \frac{(1-\epsilon-\epsilon_{\rm k})\, L_{\rm
k}}{c^2\epsilon_{\rm k}} \label{eq|SoltanKin}\, ,
\end{equation}
where $\Phi(L_{\rm k},z)$ is the \emph{kinetic} luminosity function
of AGNs in units of ${\rm Mpc^{-3}\, dex^{-1}}$ defined below.

%The kinetic efficiency \epsikin$=\epsilon$\epsi is then computed on
%adopting a radiative efficiency of $\epsilon=0.1$, a typical value
%for radiatively efficient AGNs (e.g., So{\l}tan 1982).
The value of \epsi derived in \S~\ref{sec|results} was calibrated on
the CMC sample which is representative of radio-loud and optically
bright AGNs, mostly composed of SSRs. Neglecting to lowest order
$\epsilon$ and \epsikin in the numerator of Eq.~(\ref{eq|SoltanKin})
does not alter our conclusions. We now use as our wider database the
5 GHz radio LF by De Zotti et al. (2005) specific for SSRs. The
Steep Spectrum kinetic luminosity function $\Phi(L_{\rm k},z)$ is
then derived by convolving this luminosity function with a Gaussian
with mean given by Eq.~(\ref{eq|LkinL}), where \epsi and the
dispersion $\sigma$ are given in Eq.~(\ref{eq|gkin}).

\begin{figure}
\epsscale{0.4} \plotone{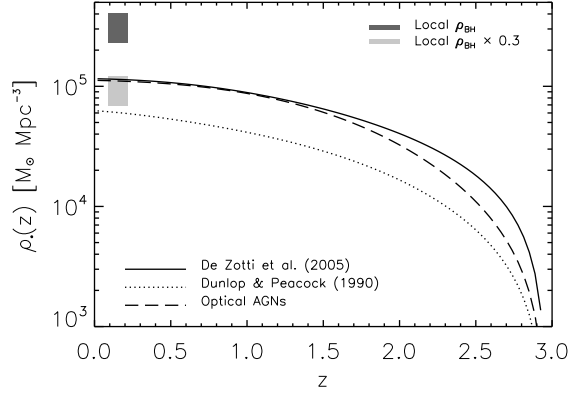}\caption{\small Cumulative black
hole mass densities computed as a function of redshift. The
\emph{long-dashed} line represents the mass density associated with
optical AGNs, obtained by integrating the Richards et al. (2005)'s
LF above $M_B=-21$ (i.e., $\log P_{1.4\, {\rm GHz}}/({\rm W/Hz/sr})
\sim 23.1$ and $\log L_{\rm k}/({\rm erg/s})\sim 43.9$, on using
eq.~(\ref{eq|gkin})). The \emph{solid} and \emph{dotted} lines are
the results of integrating above the same luminosity threshold the
LFs by De Zotti et al. (2005) and Dunlop \& Peacock (1990),
respectively. The value of $\epsilon=0.1$ and our best-fit value
\epsi=0.10 have been used for all curves. The boxes show the local
black hole mass density $\rho_{\rm BH}$ in massive spheroids
(\emph{dark grey} area) and 30\% of this value (\emph{light grey}
area).} \label{fig|rhokin}
\end{figure}

Our result is shown by the solid line in Fig.~\ref{fig|rhokin}. It
is seen that at $z\approx 0$ the BH mass density we evaluate to be
associated to the Steep Spectrum radio-loud population is about
$25\%$ of the BH mass density found in local
galaxies (light grey area). %Interestingly we find that $\rho_{\rm k}$ matches a
%value of $0.2\rho_{\rm BH}$ instead (light grey area), supporting
%the idea that only a minority of the relic BHs present in
%the local universe have undergone a radio-loud kinetic phase (at
%least the more massive ones). Our estimate for $\rho_{\rm k}$
%would instead agree with the full value of $\rho_{\rm BH}$ only on
%using $\epsilon_{\rm k}=0.002$ (dotted line), which is, however,
%about $7\, \sigma$ below our best-fit value.

This amount is not far from the cumulative mass density obtained
from the optical LF by Richards et al. (2005), representative of the
whole population of optical AGNs at $z\lesssim 3$, which is shown
with a long-dashed line in Fig.~\ref{fig|rhokin}. The latter mass
density is obtained via the standard radiative So{\l}tan argument
which is formally equivalent to Eq.~(\ref{eq|SoltanKin}) with
$\epsilon_{\rm k}$ in the denominator replaced by $\epsilon$, and
$L_{\rm k}$ by $L$. Also shown with a dotted line in
Fig.~\ref{fig|rhokin} is the BH mass density obtained by integrating
the SSR LF by Dunlop \& Peacock (1990); a slightly higher value
would be obtained by integrating the SSR LF by Willott et al.
(2001). These differences in BH mass densities simply reflect
differences in the SSR LF adopted at given kinetic efficiency.

In any event, our computations converge to show that the accreted
mass density associated with Steep Spectrum radio-loud kinetic AGNs
alone is comparable to the total BH mass density accreted by optical
AGNs. Note that the optical luminosity function has been extended
down to $M_B\sim -21$, which is the minimum luminosity probed by
Richards et al. (2005). On the other hand, our mass density
evaluations derived from the SSR LFs have been computed by extending
the computations to the corresponding limiting luminosity, i.e.
$\log L_{\rm k}/({\rm erg\, s^{-1}})\sim 43.9$ and $\log P/({\rm W\,
Hz^{-1}\, sr^{-1}})\sim 23.1$, on the basis of Eqs.~\ref{eq|Willott}
and ~\ref{eq|gkin}.

\begin{figure}
\epsscale{0.8} \plotone{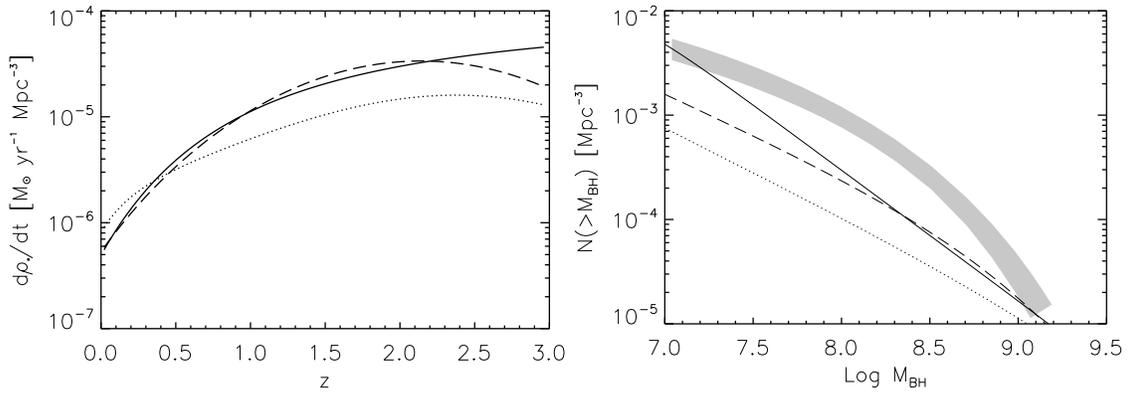}\caption{\small \emph{Left panel}:
differential accretion density as a function of redshift.
\emph{Right panel}: the corresponding cumulative number density at
$z\sim 0$ of black holes derived from the equation representing the
building-up of $\rho_{\bullet}(z)$ by gravitational interactions of
their host galaxies (see Vittorini, Shankar \& Cavaliere 2005, and
Shankar et al. 2007). The line styles in both panels have the same
meaning as in Fig.~\ref{fig|rhokin}. The \emph{grey} band shows the
cumulative local mass function with its uncertainty region taken
from Shankar et al. (2007).} \label{fig|NM}
\end{figure}

In the left panel of Fig.~\ref{fig|NM} we show the differential mass
density as a function of $z$ for radio-loud and optical AGNs. It can
be seen that the approximate match between the mass density accreted
by kinetic and optical AGNs holds at each redshift, supporting a
scenario in which the optical and kinetic outputs are produced by
similar accretion events at each time.

The match also holds in the final cumulative BH mass functions, as
shown in the right panel of Fig.~\ref{fig|NM}. The grey area in the
Figure represents the local BH mass function with its systematic
uncertainties. Keeping in mind that below a BH mass $M_{\bullet}\sim
10^8\, {\rm M_{\odot}}$ our results are actually extrapolated, it
can be seen that the fraction of the local BH population which has
been kinetically active as Steep Spectrum sources in its past grows
with $M_{\bullet}$ from about $25\%$ to reach near totality for
masses $\gtrsim 10^9\, {\rm M_{\odot}}$. This was computed on
adopting an average Eddington ratio of $\dot{m}=L/L_{\rm Edd}=0.5$,
where $L_{\rm Edd}=1.3\times 10^{38} M_{\bullet}/{\rm M_{\odot}}\,
{\rm erg\, s^{-1}}$. This ratio is found to provide a good match
between the local and accreted mass functions for the overall AGN
population by, e.g., Marconi et al. (2004), Shankar et al. (2004)
and Shankar, Weinberg \& Miralda-Escud\'{e} (2007); our results
however do not significantly depend on the precise choice for the
adopted Eddington ratio, within the range of several tenths.

The results presented in this \S$\,$ indicate a considerable
contribution to the local mass density from radio-kinetically active
SSRs, in fact comparable to the one from optical/X-ray sources
(e.g., Shankar et al. 2004). The latter ones, not explicitly
accounted for here, are able to explain the mass around
$M_{\bullet}\sim 10^8\, {\rm M_{\odot}}$ needed to attain full match
with the local mass function (see right panel of Fig.~\ref{fig|NM}).
In other words, a considerable fraction of relic black holes have
possibly undergone a kinetically-loud phase in their past. %Such
\section{DISCUSSION AND CONCLUSIONS}
\label{sec|conclu}

We have proposed a powerful method for estimating the average
efficiency of the \emph{kinetic} power production process, derived
essentially from observational constraints, i.e., from
intercalibrating the radio and the optical LFs. Our result therefore
is independent of specific assumptions on the jet or wind dynamics
and composition; it can be actually used to constrain theoretical
models of jets. The value for the constant $f\simeq 15$ discussed by
Hardcastle et al. (2007) and inserted in Eq.~(\ref{eq|Willott}),
supports the picture of "heavy" jets with a dominant protonic
component; this enhances the kinetic output associated with even
modest radio luminosities. We have adopted Eq.~(\ref{eq|Willott})
for all radio-loud sources, irrespective of their redshift and/or
environment.

%In our calculations in \S~\ref{sec|results} we have corrected the
%radio and optical LFs for incompleteness effects including the
%normalization factors derived by CMC. We are confident that our
%results should not be affected by other systematics in the
%determination of the LFs used for our computation in
%\S~\ref{sec|results}.
So far we did not consider effects of beaming in the Steep Spectrum
radio LF. We expect the degree of beaming in these sources to be
small on average. In fact, we have checked that on correcting the
observed luminosities in the CMC radio LF with the beaming
parameters appropriate for SSRs as done in the unification models of
Urry \& Padovani (1991, 1995), the resulting intrinsic LF is very
similar to the one observed by CMC, only $\approx 0.06$ dex fainter
on average. This in turn implies values about $10\%$ smaller for the
$g_{\rm k}$ best value, still within our $1-\sigma$
uncertainties.%\footnote{We here did not correct the normalization of
%the LF in this computation, as we are not interested in recovering
%the full intrinsic distribution of the parent population but only to
%correct the observed radio luminosities in our sample for effects of
%beaming.}.

%considering also the Flat Spectrum component in Steep spectrum AGNs,
%by integrating over redshift and luminosity the de-boosted Dunlop \&
%Peacock (1990)'s LFs. By dividing the cumulative kinetic energy
%density by the total local BH mass density, Heinz et al. (2007) find
%kinetic efficiencies higher by a factor of a few relative to our
%results.
Our statistical results on the average kinetic efficiency of SSRs do
not exclude that FSQs could have a higher kinetic efficiency, or
that there could be sources with exceptionally high kinetic outputs.
For example, on discussing individual powerful blazars Maraschi \&
Tavecchio (2003) find higher values for the kinetic efficiency than
derived here. This is partly due to their selection of the brightest
jets for which the SEDs can be determined up to $\gamma$-ray range.
In addition, the high energy emissions probe the jet closer to the
nucleus; conceivably the jet decelerates before reaching the radio
hot spots and lobes, so that part of its kinetic energy is lost on
the way. Thus the extended radio emissions on which
Eq.~(\ref{eq|Willott}) has been calibrated could only set a lower
limit to the full kinetic energy imparted to the jet at its origin.
Heinz et al. (2007) have recently estimated the cumulative FSQ
kinetic energy, finding average kinetic efficiencies a factor of a
few higher than our best-fit value, but still on the order of a few
percent of accreted rest-mass energy.

To make contact with these independent studies, we have redone our
calculation on adopting for the correction factor $f$ a higher value
than the already significant one used here; for example, setting
$f=45$ in Eq.~(\ref{eq|Willott}), we get a relation very similar to
that derived by Heinz et al. (2007). Our $\chi^2$ fitting yields in
this case \epsi=$0.53^{+0.20}_{-0.27}$ and
$\sigma=0.307^{+0.003}_{-0.004}$, which implies kinetic efficiencies
a factor of five higher than derived in \S~\ref{sec|results}. While
more stringent calibrations on relations such as that in
Eq.~(\ref{eq|Willott}) are essential to pin down the average value
of \epsikin, all methods support kinetic efficiencies of order a few
percent. Consider that having increased $f$ and \epsi$\,$ will
proportionally increase the kinetic efficiency and mean kinetic
luminosity, although leaving the BH mass density $\rho_{\rm
BH}\propto L_{\rm k}/\epsilon_{k}$ unaltered.

%Note that our result on \epsi agrees with the average value derived
%by Ghisellini \& Celotti (2001) for FRI and FRII radio sources, if
%the dividing line between the two populations is identified with the
%transition to subcritical accretion rates.
Finally, we note that the levels of kinetic efficiency derived in
this work agree with the amount of energy kinetic feedback required
in theoretical studies of massive galaxies (e.g., Granato et al.
2004, Cavaliere \& Lapi 2006, Shankar et al. 2006).

Summarizing, we have used an optically selected sample of radio loud
AGNs dominated by SSRs to convert the optical LF to a radio one. We
have found that the kinetic output in jets or winds must amount to a
fraction \epsi$\sim 10^{-1}$ of the bolometric luminosity to match
the empirical radio LF for the same sample; with this value we have
been able to reproduce also a number of empirical correlations
relating $L_B$, $L_{\rm k}$ and $L_R$. From typical values for the
AGN radiative efficiency $\epsilon\sim 10^{-1}$, we have derived an
average kinetic efficiency \epsikin$\sim 10^{-2}$ which we have
assumed to be common to all SSRs. By using Eq.~(\ref{eq|SoltanKin}),
we have then constrained the contribution of kinetically-loud SSRs
to the local BH mass density to be $\gtrsim 25\%$, comparable to
what is found from optical/X-ray selected AGNs. We consider the
latter estimate to provide a firm \emph{lower} limit; additional
contributions to the kinetic integrated AGN emission could come from
the more numerous and less luminous AGNs (e.g., Heinz et al. 2007),
from AGNs in radiatively inefficient but kinetically efficient
accretion states (e.g., Churazov et al. 2005), and/or from FSQs with
intrinsically higher kinetic power than here considered (see
Maraschi \& Tavecchio 2003). An additional contribution arises from
radio galaxies which should have comparable kinetic outputs
according to Unification models; in agreement with our findings in
fact, Koerding et al. (2007) find $g_{\rm k}\approx 0.2$ for a
significant sample of FRI and FRII radio galaxies.

% Finally, according to AGN
%Unification Models, all radio galaxies are radio AGNs on the plane
%of the sky and therefore should be added to the total kinetic AGN
%LF, thus possibly further increasing the predicted local demography
%of the kinetically loud BHs.
%
%We argue that the magnetic extraction of energy from BHs to be often
%subdominant with respect to the direct energy extraction from the
%accretion flow itself.
%What is interesting however is that a constant factor (\epsi)
%between optical luminosity and kinetic luminosity accounts for a
%wide range of scaling relations.

\begin{acknowledgements}
Work partly supported by NASA grant GRT000001640. We thank A.
Merloni for discussions; FS thanks S. Mathur for comments on an
earlier version of the manuscript. Thanks are due to our referee for
suggestions helpful toward improving our presentation.
\end{acknowledgements}

%;;;;;;;;;;;;;;;;;;;references;;;;;;;;;;;;;;;;;

%-----------------------------------------------------------------
%\begin{figure}
%\epsfxsize=7.5cm \epsffile {fig1.eps}
%\begin{figure}
%\epsscale{0.6} \plotone{fig1.eps} \caption{\small Comparison between
%the fraction of radio-loud AGNs in optical samples at two redshifts
%as labeled. From the Richards et al. (2005) luminosity function we
%extract the expected number of radio loud sources as given by Jiang
%et al. (2007); the results is shown with a \emph{dashed} line with
%its uncertainty region (striped area). The filled circles are the
%data by Cirasuolo et al. (2005) renormalized by a factor of 1.5 to
%account for incompleteness, while the \emph{solid} line is their
%best-fit luminosity function with its uncertainty region
%(\emph{grey} area).} \label{fig|duty}
%\end{figure}
%
%
%\begin{figure}
%\epsfxsize=7.5cm \epsffile {fig4.eps}
%
%\begin{figure}
%\epsfxsize=7.5cm \epsffile {fig2.eps}
\newpage

\newpage
%\begin{figure}
%\epsfxsize=7.5cm \epsffile {fig3.eps}

\newpage

\newpage

\newpage
%
%\begin{figure}
%\epsscale{0.6} \plotone{fig5_1.eps}\caption{\small Parameter plane
%for $f_{\rm k}$ as a function of $\epsilon$ and $g$. The
%\emph{striped} area is the one where $f_{\rm k}<0$. The
%\emph{grey} areas show the regions not allowed by the constraints on
%the radiative efficiency. The lower solid lines are contour levels
%for $f_{\rm k}$, with values as labeled. A fixed value of
%\epsi=0.15 has been used.} \label{fig|fkin}
%\end{figure}

\newpage

%-----------------------------------------------------------------

%\end{doublespace}

\end{document}